\documentclass[10pt]{iopart}

\usepackage[ansinew]{inputenc}
\usepackage{graphicx}
\usepackage{siunitx}
\usepackage{nicefrac}

\newcommand{\eg}{{\em e.\,g.}}
\newcommand{\ie}{{\em i.\,e.}}

\begin{document}

\title[The Kondo Resonance Line Shape in STS]{The Kondo Resonance Line Shape in Scanning Tunnelling Spectroscopy: Instrumental Aspects}

\author{Manuel Gruber$^{1}$, Alexander Weismann$^{1}$, Richard Berndt$^{1}$}
\address{$^1$Institut f{\"u}r Experimentelle und Angewandte Physik, Christian-Albrechts-Universit{\"a}t zu Kiel, D-24098 Kiel, Germany}
\ead{gruber@physik.uni-kiel.de}

\begin{abstract}
In the scanning tunnelling microscope, the many-body Kondo effect leads to a zero-bias feature of the differential conductance spectra of magnetic adsorbates on surfaces.
The intrinsic line shape of this Kondo resonance and its temperature dependence in principle contain valuable information.
We use measurements on a molecular Kondo system, all-\textit{trans} retinoic acid on Au(111), and model calculations to discuss the role of instrumental broadening.
The modulation voltage used for the lock-in detection, noise on the sample voltage, and the temperature of the microscope tip are considered.
These sources of broadening affect the apparent line shapes and render difficult a determination of the intrinsic line width, in particular when variable temperatures are involved.
\end{abstract}


\maketitle
\ioptwocol

\section{Introduction}

The Kondo effect arises from the interaction of a localized spin with conduction electrons of a host metal \cite{kondo_resistance_1964,hewson_kondo_1993}.
Using scanning tunnelling spectroscopy (STS), one of its fingerprints, a resonance close to the Fermi energy $E_F$, \ie\ at zero bias voltage, has been investigated for metal adatoms \cite{li_kondo_1998, madhavan_tunneling_1998, nagaoka_temperature_2002, knorr_kondo_2002, wahl_kondo_2004, limot_kondo_2004, choi_conductanceDriven_2012, von_bergmann_spin_2015} and adsorbed molecules \cite{wahl_kondo_2005, zhao_controlling_2005, iancu_manipulation_2006, iancu_manipulating_2006, gao_siteSpecific_2007, fernandezTorrente_vibrational_2008, perera_spatially_2010, choi_single_2010, mugarza_spin_2011, komeda_observation_2011, tsukahara_evolution_2011, franke_competition_2011, dilullo_molecular_2012, robles_spin_2012, gopakumar_electronInduced_2012, miyamachi_robust_2012, minamitani_symmetryDriven_2012, strozecka_reversible_2012, kim_switching_2013, heinrich_change_2013, lin_twoDimensional_2014, minamitani_spatially_2015, zhang_kondo_2015, wu_modulation_2015, esat_transfering_2015, karan_shifting_2015, esat_chemically_2016, meyer_influence_2016, warner_sub-molecular_2016, karan_generation_2016, karan_spin_2016, ormaza_surface_2016, knaak_ligandInduced_2017, knaak_interconnected_2017, li_anomalous_2017, pacchioni_two-orbital_2017, gruber_spin_2017, zhang_2017, hiraoka_singlemolecule_2017, ormaza_controlled_2017}.
Recent reviews are available \cite{ternes_spectroscopic_2008, ternes_probing_2017}.
STS offers the advantage that both occupied and unoccupied states close to $E_F$ can be sensitively probed.
Moreover, the high spatial resolution of STS can be used to, \eg, locate different Kondo resonances within a single molecule \cite{pacchioni_two-orbital_2017, knaak_ligandInduced_2017}.
The STS line shape of a Kondo resonance is not trivial as it depends on (i) the intrinsic properties the Kondo system, which in turn are affected by temperature and magnetic fields, (ii) matrix element effects of the tunnelling process, which may be interpreted in terms of an interference between different tunnelling channels, and (iii) instrumental broadening.
The splitting or the broadening of the Kondo resonance in magnetic fields and at elevated temperatures, respectively, have sometimes been used to exclude other effects that may lead to zero-bias features in STS\@.
Initially, STS data of Kondo systems were fitted with Fano line shapes \cite{fano_effects_1961}: 
\begin{equation}
Fano (V) \propto \frac{\left( q + \epsilon \right)^2}{1 + \epsilon ^2}
\label{eq:Fano}
\end{equation}
\noindent where $V$ is the sample voltage and $\epsilon$ is a normalized energy, expressed as 
$ \epsilon = (eV - E_K)/\Gamma_\text{Fano}.$
\noindent $E_K$ and $\Gamma_\text{Fano}$ are the energy and the half-width at half-maximum (HWHM) of the resonance, respectively.
The asymmetry factor $q$ varies the line shape of the resonance from a Lorentzian peak ($q \rightarrow \infty$) over a asymmetric feature  to a dip ($q \rightarrow 0$).
It reflects the relative importance of tunnelling to the localized resonance and the delocalized band electrons.

In 1992, Frota had suggested a different line shape to fit Kondo resonances \cite{frota_shape_1992} but it was first employed only some 20 years later for STS measurements \cite{pruser_longrange_2011, pruser_mapping_2012}.
The Frota line shape is a fit to the shape of the Kondo resonance found from numerical renormalization group calculations.
It may be expressed as follows \cite{pruser_mapping_2012}:
\begin{equation}
Frota (V) \propto \Im \left[ i e^{ i \phi } \sqrt{\frac{i \Gamma_\text{Frota}}{eV - E_K + i \Gamma_\text{Frota}}} \right]
\label{eq:Frota}
\end{equation}
\noindent $\phi$ is a form factor playing a similar role as $q$ for the Fano line shape.
The HWHM of the line is given by $2.542\,\Gamma_\text{Frota}$ \cite{pruser_longrange_2011}.

The characteristic energy scale of the Kondo effect is usually expressed via the Kondo temperature $T_K$.
In principle, $T_K$ is directly related to the width of the spectral feature in STS\@.
However, other sources of broadening must be properly taken into account.

Here, we present experimental and modelling results on the influence of voltage noise and the temperature of measurement on the Kondo line shape.
For the experiments, we used retinoic acid (ReA) on Au(111) \cite{karan_spin_2016, karan_generation_2016, gruber_manipulation_2016, zhang_2017} because (i) the magnetic origin of the resonance has been proven with measurements in magnetic fields, (ii) the differential conductance spectrum exhibit no other features close to the Kondo resonance (\eg, no $d$ orbital is overlapping the resonance), and (iii) Kondo resonances of different widths can be obtained on the same system.

We find that a Frota line shape fits best to the experimental data, provided that voltage noise has been sufficiently reduced.
Voltage variations caused by, \eg, a modulation used for lock-in detection or environmental sources, broaden and deform the resonance.
Because of these deformations, a Fano line shape may actually become a better match of the noise-affected data.
Temperature broadening is predicted to have a similar effect.
We describe the instrument functions that in principle may be used to deconvolve the data.
However, an accurate determination of line widths remains challenging.

\section{Methods}

The experiments were performed in ultra-high vacuum using a scanning tunnelling microscope operated at \SI{4.4}{\kelvin} (Createc, Berlin).
Tungsten tips were treated \textit{in situ} by field emission followed by indentation into the Au(111) substrate.
Au(111) single-crystal surfaces were prepared by cycles of Ar-ion bombardment and annealing to \SI{860}{\kelvin}.
All-\textit{trans} ReA molecules (Sigma-Aldrich) were thermally sublimated at a pressure of \SI[exponent-product = \times]{\approx 1e-9}{\milli\bar}.

The differential conductance of the tunnelling junction was measured using a lock-in amplifier.
A modulation with an amplitude $V_m =\SI{420}{\micro\volt}$ at a frequency of \SI{737.4}{\hertz} was applied to the sample voltage $V$ unless otherwise specified.
The spectra were acquired with a feedback loop opened at a current of \SI{100}{\pico\ampere} and a voltage of 50 or \SI{100}{\milli\volt}.
The spectra shown are averages over 4 voltage sweeps.
In addition, the data at large $|V|$ beyond the resonance were low-pass filtered.

In addition to $V_m$, further sinusoidal modulations ($V_1$) were added to the sample bias. 
These modulations were generated using an external function generator (Agilent 33220A) and internal generators of the STM control electronics (Specs Nanonis).
They were disconnected by relays if unused.
A home-made circuit was used for summation and voltage amplitude reduction (factor 100). 
When not otherwise specified, the sample bias signal was transmitted through an LC low-pass (cutoff \SI{10}{\kilo\hertz}) and a feedthrough $\Pi$ filter (cutoff \SI{1}{\mega\hertz}).

\section{Results and discussion}

\begin{figure}
\begin{center}
\includegraphics{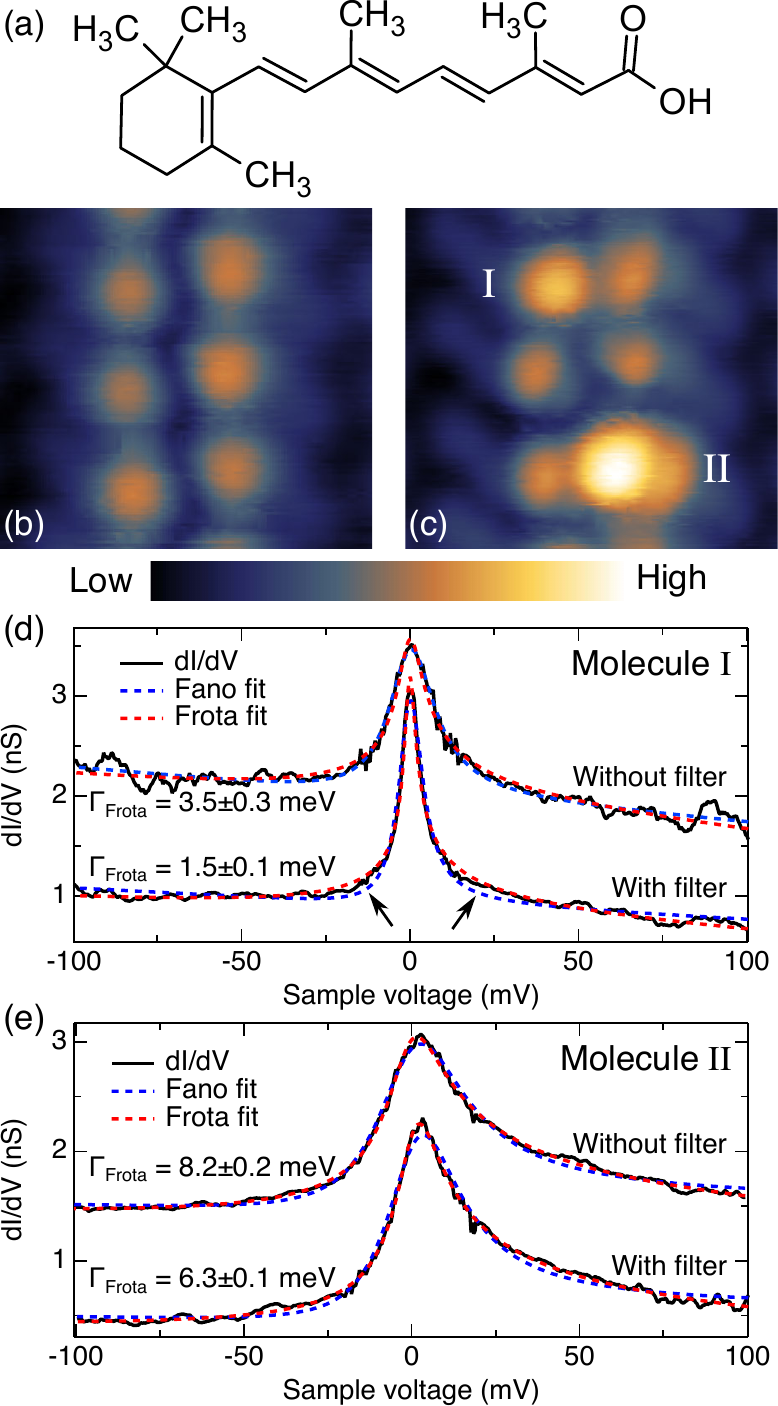}
\end{center}
\caption{(a) Scheme of all-\textit{trans} retinoic acid (ReA).
(b) and (c) STM topographs of six ReA molecules on Au(111) before and after switching the molecules marked I and II\@.
Widths of the STM topographs \SI{6}{\nano\meter}.
Colour scales range (a) from 0 to \SI{200}{\pico\meter} ($I = \SI{30}{\pico\ampere}$ and $V = \SI{1.0}{\volt}$) and (b) from 0 to \SI{280}{\pico\meter} ($I = \SI{30}{\pico\ampere}$ and $V = \SI{0.1}{\volt}$).
(d) and (e) $dI/dV$ spectra (black) of molecules I and II acquired with (lower curve) and without (upper curve) a low-pass filter in the voltage line, along with Fano (dashed blue) and Frota (dashed red) fits.
All Fano and Frota fits allowed for a linear background. 
Arrows indicate maximum deviations of the Fano fits from the data.
The spectra recorded without filters are shifted by \SI{1}{\nano\siemens} for clarity.} 
\label{fig:topoAndSpectraOverview}
\end{figure}

ReA molecules on Au(111) arrange into an ordered pattern in islands \cite{karan_spin_2016, karan_generation_2016, gruber_manipulation_2016, zhang_2017}.
Figure~\ref{fig:topoAndSpectraOverview}b shows a typical topograph of six ReA molecules on Au(111).
Each ReA molecule exhibits a round protrusion at the expected location of the cyclohexene group.
Differential conductance ($dI/dV$) spectra of the pristine molecules show no feature close to $V=0$ (corresponding to $E_F$) \cite{karan_spin_2016}.
Upon application of a voltage of \SI{-2.2}{\volt} for a few seconds ($I = \SI{100}{\pico\ampere}$), the molecule under the tip may be switched into different states.
Figure~\ref{fig:topoAndSpectraOverview}c shows a STM topograph upon two successful switching events (same scanned area as in Fig.~\ref{fig:topoAndSpectraOverview}b).
The molecules marked I and II appear higher than the reference molecules nearby.
Their $dI/dV$ spectra exhibit sharp peaks at the Fermi level (Figs.~\ref{fig:topoAndSpectraOverview}d--e) owing to a Kondo resonance as verified by measurements in magnetic fields \cite{karan_spin_2016}.

Next, we present the effects of voltage noise and the voltage modulation amplitude on the line shapes.
Figure~\ref{fig:topoAndSpectraOverview}d shows two $dI/dV$ spectra acquired over molecule I, along with fits of Fano (Eq.~\ref{eq:Fano}) and Frota (Eq.~\ref{eq:Frota}) functions.
The spectra were recorded with (lower curve) and without (upper curve) low-pass filters in the voltage line, mounted as close as possible to the STM\@.
The widths of the peak around $E_F$ are strikingly different as is evident from the respective values of $\Gamma_\text{Frota}$ (1.5 and \SI{3.5}{\milli\electronvolt}).
While it may not come as a surprise that line widths are strongly affected by instrumental noise the spectra highlight that voltage noise has to be independently characterized to obtain reliable estimates of the Kondo temperatures.
To characterize our instrument, $dI/dV$ spectra were acquired with low-pass filters on Pb(111).
Fits of the measured superconducting gap lead to an effective temperature $T \approx \SI{5}{\kelvin}$ indicating that the broadening is limited by temperature.

A second difference between the spectra in Fig.~\ref{fig:topoAndSpectraOverview}d is more subtle but equally important.
The voltage noise affects the quality of the Frota and Fano fits defined by the coefficient of determination $R^2$:
\begin{equation}
R^2 = 1 - \frac{\sum_i (y_i - f_i )^2}{ \sum_i (y_i - \bar{y})^2},
\label{eq:Rsquare}
\end{equation}
\noindent where $y_i$ is $i^\text{th}$ measured $dI/dV$ value, $\bar{y}$ the average of the $dI/dV$ data, and $f_i$ is the value of the fit function for the point $i$.
$R^2$ approaches one as the fit improves.
The high-quality data obtained with a filter are better fit by a Frota line ($R^2_\text{Frota} = \num{0.985} > R^2_\text{Fano} = \num{0.974}$).
For the noise-affected data, the situation is reversed: $R^2_\text{Frota} = \num{0.957} < \num{0.965} = R^2_\text{Fano}$.
Below we use numerical simulations to show that this trend --a better fit of Fano lines in the presence of noise-- is rather general.

Related measurements were performed on molecule II (Fig.~\ref{fig:topoAndSpectraOverview}e).
Because of the larger intrinsic width of the Kondo resonance in this case, the impact of the instrumental broadening is less obvious and Frota lines better match the data with and without filter.
Nevertheless, the instrumental noise increases the peak width by approximately \SI{2}{\milli\electronvolt}, similar to the case of molecule I\@.

\subsection{Broadening introduced via the sample voltage}

\subsubsection*{Voltage modulation for lock-in detection\\[3mm]}

$dI/dV$ spectra are usually acquired using a lock-in amplifier to improve the signal-to-noise ratio.
This technique requires the addition of a (usually) sinusoidal modulation voltage (amplitude $V_m$) to the sample voltage and consequently introduces some broadening.

\begin{figure}
\begin{center}
\includegraphics{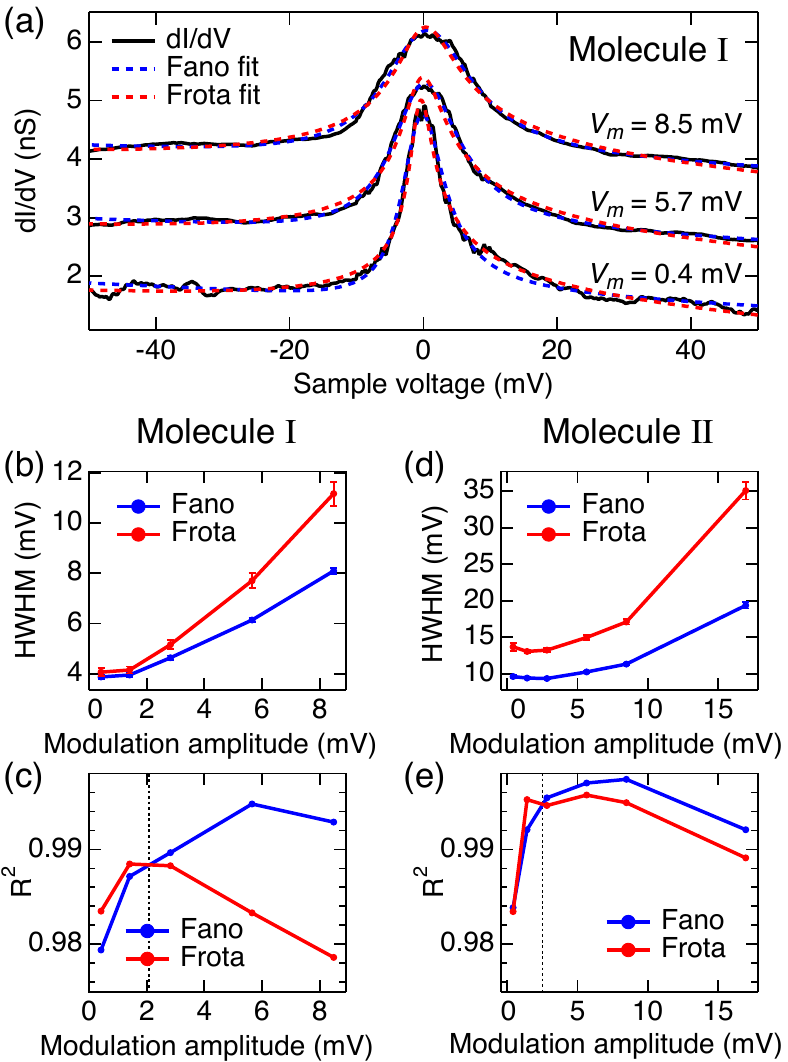}
\end{center}
\caption{(a) $dI/dV$ spectra (black) acquired over molecule I with modulation amplitudes $V_m$ of 0.4, 5.7, and \SI{8.5}{\milli\volt}, along with Fano (dashed blue) and Frota (dashed red) fits.
(b) and (c) Evolution of the HWHM extracted from Fano (blue) and Frota (red) fits as a function of the modulation amplitude for molecules I and II\@.
(d) and (e) $R^2$ vs.\ $V_m$ for molecules I and II\@.} 
\label{fig:lockinBroadening}
\end{figure}

Spectra of molecule I acquired with different modulation amplitudes are presented in Fig.~\ref{fig:lockinBroadening}a.
The HWHM was extracted from Fano and Frota fits to the spectra, using the relations $\text{HWHM} = \Gamma_\text{Fano}$ and $\text{HWHM} = 2.542\,\Gamma_\text{Frota}$, respectively.
The resulting evolution of the resonance width is displayed in Fig.~\ref{fig:lockinBroadening}b.
While modulations of \SI{0.4}{\milli\volt} to \SI{1.4}{\milli\volt} have little impact, larger $V_m$ cause an approximately linear increase of the HWHM\@.
It is worth mentioning that the widths extracted from a Fano and Frota fits are different because none of these line shapes perfectly matches the data.
In other words, the error made by neglecting the broadening depends on the line shape used.

Figure~\ref{fig:lockinBroadening}c shows $R^2$ vs.\ the modulation amplitude for Fano (blue) and Frota (red) fits.
All fits seem very good ($R^2$ of exceeding 0.98) and do not indicate the presence of systematic errors.
As expected from the discussion above, $R^2$ of the Fano fits further improves as the broadening due to $V_m$ is increased.
For $V_m < \SI{2}{\milli\volt}$ ($V_m > \SI{2}{\milli\volt}$) the data are better fitted with a Frota (Fano) function.
For very large modulation amplitudes, the quality of the fits is reduced.

Similar results were obtained on molecule II as shown in Figs.~\ref{fig:lockinBroadening}d--e.
The HWHMs extracted from the fits evolve slowly up to \SI{8.5}{\milli\volt} and then increase more rapidly.
The intrinsically broader Kondo resonance in this case is more robust against modulation-induced broadening.
Again, Fano fits seem superior at large modulation voltages.

\begin{figure}
\begin{center}
\includegraphics{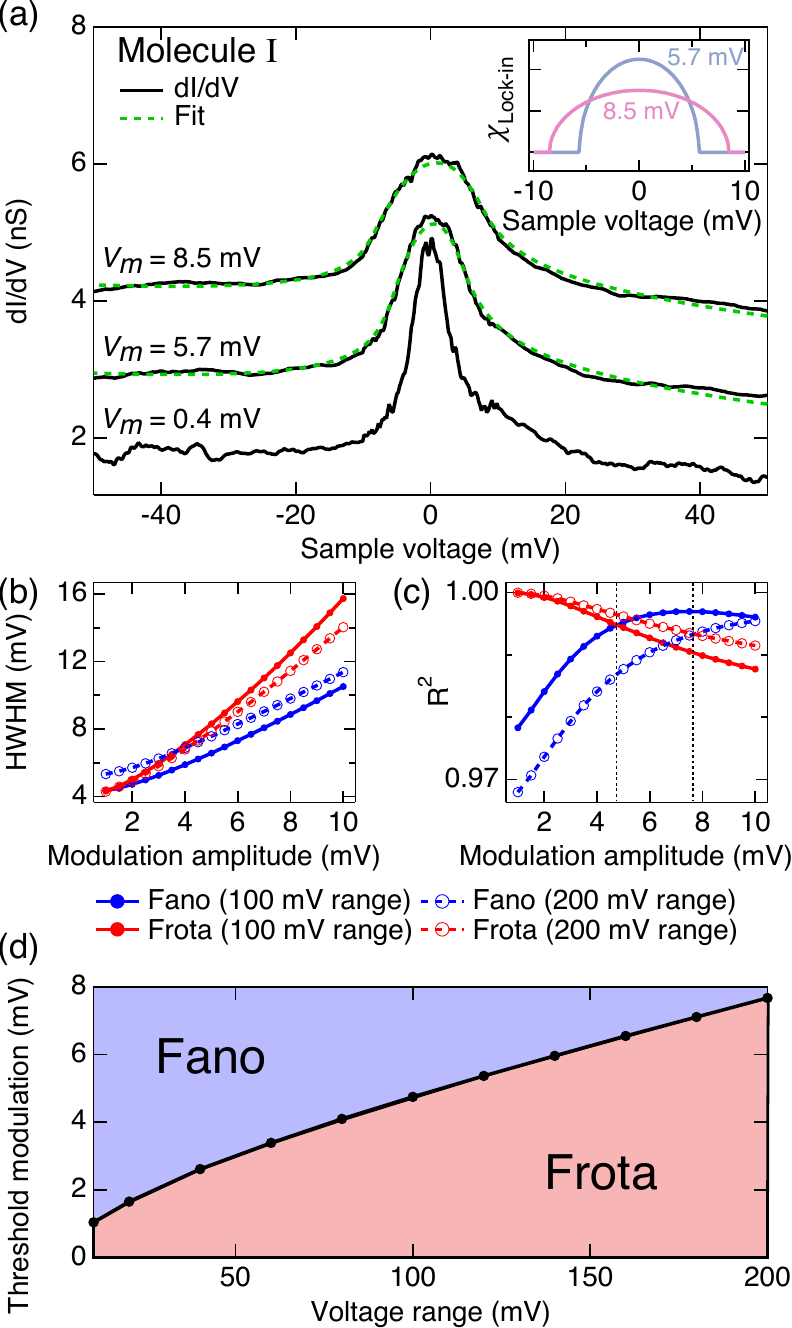}
\end{center}
\caption{(a) $dI/dV$ spectra acquired over molecule I with different modulation amplitudes $V_m$ (solid black curve).
The dashed green curves are obtained from a convolution of the $V_m = \SI{0.4}{\milli\volt}$ spectrum with the corresponding lock-in broadening functions $\chi_\text{Lock-in}$ shown in the inset.
The amplitude and offset of the green curves have been arbitrarily adjusted.
(b) and (c) Simulated evolutions of HWHM and $R^2$ extracted from Fano (blue) and Frota (red) fits as a function of $V_m$.
Range of \SI{\pm 50}{\milli\volt} (solid) and \SI{\pm 100}{\milli\volt} (dashed) were used for the fits.
Dotted black lines in (c) indicate the threshold modulation amplitude, for which a Fano fit become superior to a Frota fit.
(d) Simulated threshold modulation amplitude (\ie, amplitude where Fano and Frota $R^2$ curves cross) as a function of the voltage range used for the fits.
In the parameter area coloured blue (red) a Fano (Frota) fit matches the data better.} 
\label{fig:lockinBroadeningSimulation}
\end{figure}

The broadening introduced by a modulation can be described \cite{kroger_surface_2005} by convolving the intrinsic $dI/dV$ (not affected by modulations etc.) with
\begin{equation}
\chi_\text{Lock-in} (V) = \left\{
  \begin{array}{@{}ll@{}}
    \frac{2 \sqrt{V_m^2 - V^2}}{\pi V_m^2}, & \text{if}\ |V| \leq V_m \\
    0, & \text{otherwise.}
  \end{array}\right.
\end{equation}
Two examples of $\chi_\text{Lock-in}$ are shown in the inset to Fig.~\ref{fig:lockinBroadeningSimulation}a.
Convolution of the $dI/dV$ spectrum acquired using $V_m = \SI{0.4}{\milli\volt}$ with $\chi_\text{Lock-in}$ leads to the dashed green curves (Fig.~\ref{fig:lockinBroadeningSimulation}a).
They reproduce the line shapes acquired with $V_m = 5.7$ and \SI{8.5}{\milli\volt} very well showing that the numerical convolution accurately describes the instrumental broadening.
This will be used next to demonstrate that the voltage range considered for fitting has an impact on fit results.

The best Frota fit of the spectrum recorded with $V_m = \SI{0.4}{\milli\volt}$ (Fig.~\ref{fig:lockinBroadeningSimulation}a, bottom curve) was used as a starting point.
These numerical data enable extending the fit over a wider voltage range without adding complications due to, \eg, additional spectral structures.
The HWHMs and the $R^2$ extracted from Fano (blue) and Frota (red) fits are shown in Figs.~\ref{fig:lockinBroadeningSimulation}b--c.
It turns out that the HWHMs and $R^2$ are both sensitive to the voltage range used.
For instance, the HWHMs extracted from Fano fits are systematically larger when a wider voltage range is used (open and filled blue circles).
This in turn changes the value of $V_m$ where the $R^2$ values of Fano and Frota fits cross.
Figure~\ref{fig:lockinBroadeningSimulation}d displays the evolution of this threshold modulation amplitude as a function of the voltage range used for fitting.
It reflects the different decays of Fano and Frota lines as a function of energy ($1/E$ vs.\ $1/\sqrt{E}$). 
The figure emphasizes that excessive modulation amplitudes and limited fit ranges misleadingly favour Fano line shapes.
While these pitfalls can fairly easily be circumvented, noise present on the sample voltage $V$ may be harder to avoid.
Its impact is analysed below.

\subsubsection*{Noise of the sample voltage\\[3mm]}

\begin{figure}
\begin{center}
\includegraphics{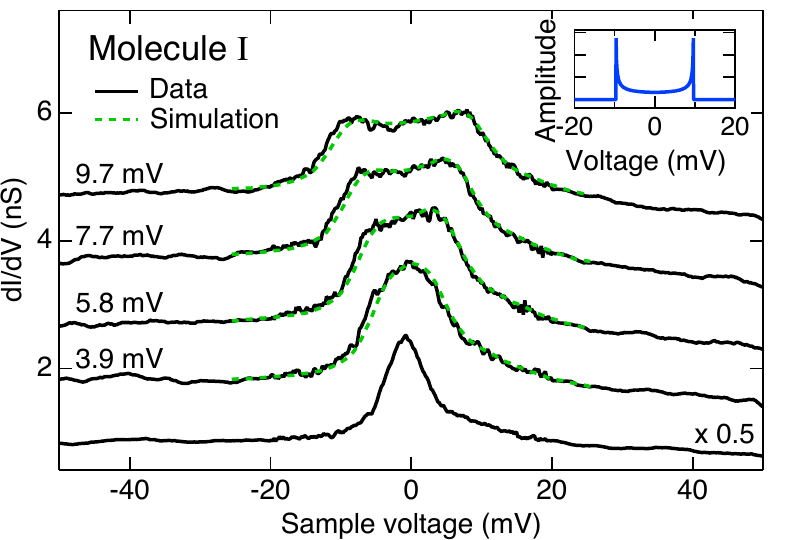}
\end{center}
\caption{$dI/dV$ spectra acquired over molecule I with sine waves at a frequency of \SI{150}{\hertz} and different amplitudes $V_1$ added to the sample voltage.
The inset shows a voltage histogram (arcsine distribution) for a sine wave with $V_m,=\SI{9.7}{\milli\volt}$.
The green curves were obtained by convolving the lower $dI/dV$ spectrum ($V_1 = 0$) with the corresponding arcsine distributions. 
For clarity, the lower spectrum has been divided by 2 and the three upper spectra have been vertically shifted by 1, 2, and \SI{3}{\nano\siemens}.
As the introduction of an external voltage effectively reduces the lock-in signal, the modulation amplitude was increased from \SI{420}{\micro\volt} to \SI{1.3}{\milli\volt}.} 
\label{fig:noise1SineWave}
\end{figure}

It is instructive to first study the effect of a single, sinusoidal signal with amplitude $V_1$ and inverse frequency $T_1$ added to $V$.
As $V_1$ is increased (Fig.~\ref{fig:noise1SineWave}), the Kondo resonance initially broadens and finally splits into two peaks.
This signal shape directly reflects the principle of measurement of the lock-in amplifier.
The differential conductance of the tunnelling gap at a time $t$, $\nicefrac{dI}{dV} \left( V + V_1 \sin \omega_1 t \right)$, is low-pass filtered, such that the  effectively measured signal is:
\begin{equation}
\overline{\frac{dI}{dV}} = \frac{2}{T_1} \int_{-T_1/4}^{{T_1/4}} \frac{dI}{dV} \left( V + V_1 \sin ( \omega_1 t) \right) dt.
\label{eq:didvAverage}
\end{equation}
Equation~\ref{eq:didvAverage} may be rewritten as: 
\begin{equation}
\overline{\frac{dI}{dV}} = \left( \frac{dI}{dV}  \ast w_1 \right) \left( V \right),
\label{eq:didvConvW}
\end{equation}
\noindent where $w_1(V)$ is the arcsine distribution \cite{paul_generation_2016}, defined as:
\begin{equation}
  w_1(V)=\left\{
  \begin{array}{@{}ll@{}}
    \frac{1}{\pi V_1} \frac{1}{\sqrt{1 - \left( \frac{V}{V_1} \right) ^2}}, & \text{if}\ |V| < V_1 \\
    0, & \text{otherwise.}
  \end{array}\right.
  \label{eq:arcsineDist}
\end{equation} 
\noindent $w_1$ is shown in the inset to Fig.~\ref{fig:noise1SineWave} using $V_1 = \SI{9.7}{\milli\volt}$.
The green curves in Fig.~\ref{fig:noise1SineWave} were obtained from Eq.~\ref{eq:didvConvW}, \ie, by convolving the $dI/dV$ spectrum, acquired without the external sine wave (lower $dI/dV$ spectrum in Fig.~\ref{fig:noise1SineWave}), with the corresponding arcsine distribution (Eq.~\ref{eq:arcsineDist}).
The match to the experimental data (black) is good as expected.
It may be noted that a similar approach is employed for sine waves in the GHz frequency range.
In that case the I-V converter is effectively averaging the current over time (and hence the differential conductance), while in the present case the averaging is performed by the lock-in amplifier \cite{paul_generation_2016,baumann_electron_2015}.

\begin{figure}
\begin{center}
\includegraphics{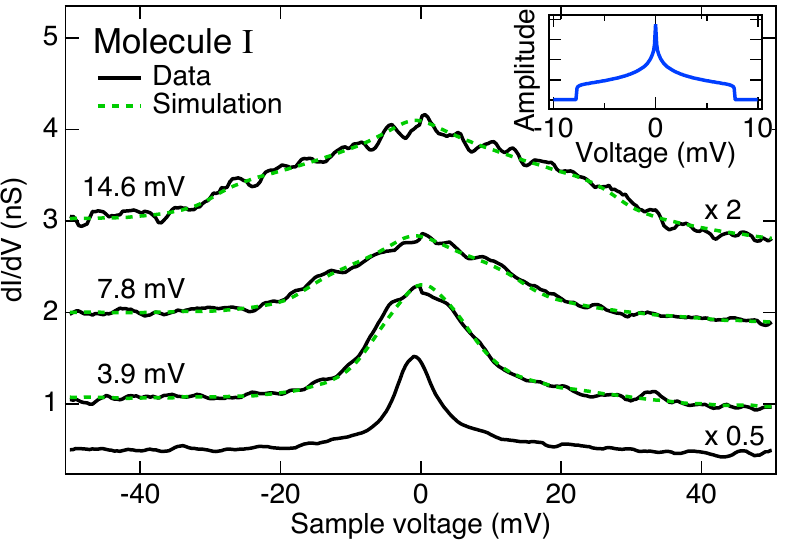}
\end{center}
\caption{$dI/dV$ spectra acquired over molecule I with sine waves at \SI{150}{\hertz} and \SI{233}{\hertz}, both of amplitude $V_1$ indicated for each curve, added to $V$.
The lower spectrum shows reference data ($V_1=0$).
The inset displays the broadening distribution ($w_1^{\ast 2} = w_1 \ast w_1$) for $V_1 = \SI{3.9}{\milli\volt}$.
The green curves were obtained by convolving the reference spectrum with the corresponding broadening distribution $w_1^{\ast 2}$. 
For clarity, the lower (upper) spectrum has been multiplied by 0.5 (2).
The two upper spectra are vertically shifted by 1 and \SI{2}{\nano\siemens}, respectively.
The modulation amplitude for lock-in detection was \SI{1.3}{\milli\volt}.} 
\label{fig:noise2SineWaves}
\end{figure}

Figure~\ref{fig:noise2SineWaves} displays experimental data on the effect of two sinusoidal modulations, both with amplitude $V_1$.
Repeating the convolution procedure described above, the measured $dI/dV$ reads:
\begin{equation}
\overline{\frac{dI}{dV}} = \left( \frac{dI}{dV}  \ast w_1 \ast w_1 \right) \left( V \right) = \left( \frac{dI}{dV}  \ast w_1^{\ast 2} \right) \left( V \right).
\label{eq:didvConvWW}
\end{equation}
\noindent 
$w_1^{\ast 2}$ is shown in the inset to Fig.~\ref{fig:noise2SineWaves} for the case $V_1 = \SI{3.9}{\milli\volt}$.
Its shape may be qualitative understood from the beating of two sinusoidal signals. 
The convolution of the lower $dI/dV$ spectrum and the corresponding $w_1^{\ast 2}$ functions (Fig.~\ref{fig:noise2SineWaves}, green lines) matches the experimental data.

\begin{figure}
\begin{center}
\includegraphics{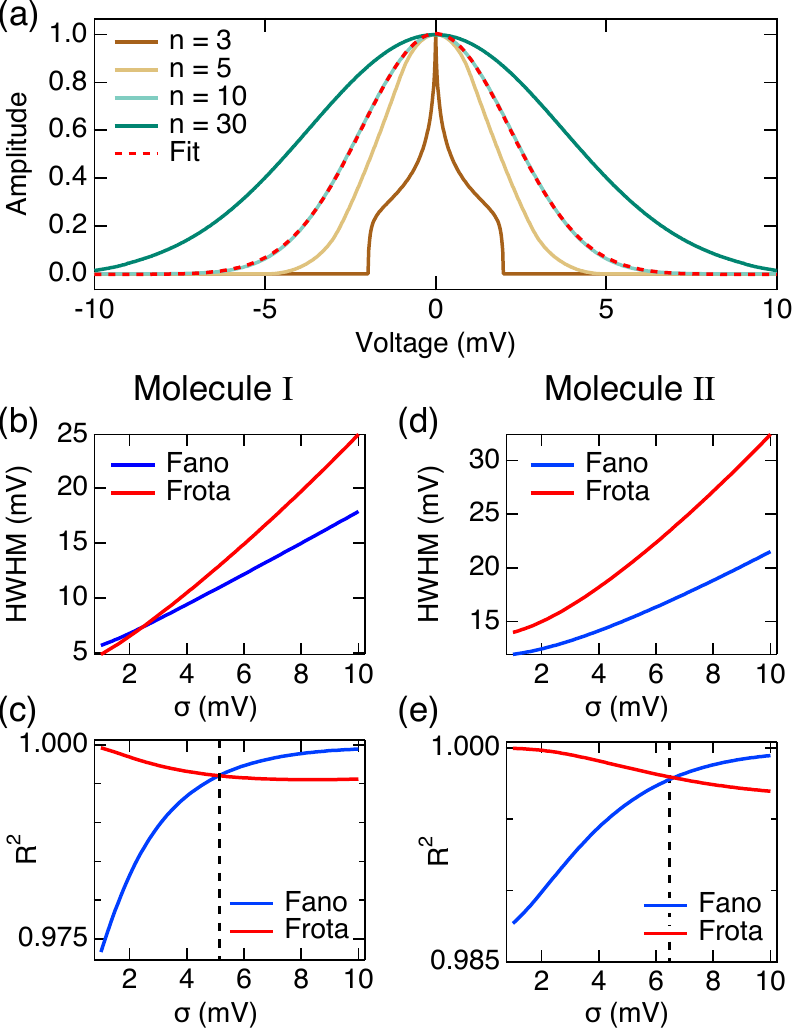}
\end{center}
\caption{(a) Broadening functions $w_1^{\ast n}$ for $n$ sine waves of amplitude $V_1 = \SI{1}{\milli\volt}$.
Maxima are normalized to 1 for clarity.
The dashed red line is a gaussian fit to $w_1^{\ast 10}$.
Gaussian fits to $w_1^{\ast n}$ lead to a standard deviation of 1.6, 2.2, and \SI{3.7}{\milli\volt} for $n = 5, 10,$ and 30\@.
The best Frota fits to the $dI/dV$ spectra on molecules I and II were convolved with a gaussian with a standard deviation $\sigma$.
(b, d) HWHMs and (c, e) $R^2$ extracted from Fano (blue) and Frota (red) fits to the convolved curves.
The $R^2$ curves intersect at $\sigma = 5.1$ and $\SI{6.6}{\milli\volt}$ for molecules I and II, respectively.} 
\label{fig:gaussianBroadeningSimulation}
\end{figure}

The procedure is easily generalized to $n$ uncorrelated sine waves of equal amplitudes $V_1$:
\begin{equation}
\underbrace{w \ast w \ast \dots \ast w}_{n} \stackrel{\text{def}}{=} w^{\ast n}.
\label{eq:convDef}
\end{equation}
As shown in Fig.~\ref{fig:gaussianBroadeningSimulation}a, $w^{\ast n}$ rapidly adopts the shape of a Gaussian, 
$ g(V; \sigma) = \frac{1}{\sigma \sqrt{2 \pi}} \exp \left( - \frac{V^2}{2 \sigma^2} \right)$.
Gaussian fits of the broadening functions lead to standard deviations 1.6, 2.2, and \SI{3.7}{\milli\volt} for $n =5, 10,$ and 30, respectively (fit for $n=5$ shown in Fig.~\ref{fig:gaussianBroadeningSimulation}a).
Note that the same standard deviations $\sigma$ can be obtained by using, for instance, a smaller amplitude $V_1$ and more sine waves.
Because of the remarkable quality of the fits, we have replaced multiple convolutions with $w$ by a single convolution with a Gaussian in what follows:
\begin{equation}
\overline{\frac{dI}{dV}} = \left( \frac{dI}{dV}  \ast g(\sigma) \right) \left( V \right).
\label{eq:didvConvGaussian}
\end{equation}

Using Eq.~\ref{eq:didvConvGaussian} we obtain the effect of an ensemble of sine waves, characterized by $\sigma$, on the line shape of a Kondo resonance.
As above, the best Frota fit of the experimental data is used as reference spectrum.
Figures~\ref{fig:gaussianBroadeningSimulation}b and c show the HWHMs and $R^2$ extracted from Fano (blue) and Frota (red) fits for molecule I as the standard deviation $\sigma$ of the gaussian is increased.
The broadening tends to favour a Fano line shape, as observed above for excessive modulation voltages $V_m$. 
Similar results for molecule II (not shown) confirm this statement.
For an ensemble of sine waves characterized by $\sigma$, we find that a narrow Kondo resonance is more strongly affected by broadening as expected.

Equation~\ref{eq:didvConvGaussian} is a simple model of the effect of broadband noise.
It may be used to further characterise the noise reduction achieved by the low-pass and $\Pi$ filters in the measurements of Fig.~\ref{fig:topoAndSpectraOverview}.
By fitting the convolution of the filtered spectra with a gaussian to the data recorded without filter we find consistent values $\sigma =  4.47$ and \SI{4.27}{\milli\volt} for molecules I and II\@.
This result may be related to a noise spectral density as follows.
White noise has a normal distribution of variance
\begin{equation}
\sigma^2 = \int_{0}^{+\infty} N_0 \, | H(f) |^2 df
\label{eq:sigmaNoise}
\end{equation}
\noindent with $N_0$ the noise power spectral density and $H(f)$ the transfer function of the system (cabling from the control electronics to the sample).
Assuming a simplified box-shaped transfer function $H(f)=1$ that drops to zero at a bandwidth $B$, we arrive at a voltage spectral density
\begin{equation}
\sqrt{N_0} = \sigma / \sqrt{B}.
\end{equation}
Using $B \approx \SI{150}{\mega\hertz}$,  $\sqrt{N_0} \approx \SI{360}{\nano\volt\raiseto{-1/2}\hertz}$ is obtained.
This value due to pick up exceeds the thermal noise and that of the control electronics \cite{burtzlaff_shot_2015} by far.

\subsection{Temperature broadening}

Smearing of the Fermi-Dirac distribution at non-zero temperature also leads to broadening of $dI/dV$ spectra.
Assuming a featureless electronic structure of the tip, the broadening can be analytically calculated, starting from the following simplified expression for the current:
\begin{equation}
I(V) = C \int_{-\infty}^{+\infty} n(E) \left[  f_\text{t} ( E-V) - f_\text{s}(E) \right] dE,
\end{equation} 
\noindent where $C$ is a constant, $n(E)$ is the sample density of states, and $f_\text{t}(E)$ and $f_\text{s}(E)$ are the Fermi-Dirac distributions of the tip and the sample:
\begin{equation}
f_\text{t/s}(E) = \frac{1}{1 + e^{-\beta_\text{t/s} E}}.
\end{equation}
\noindent $\beta_\text{t/s} = ( k_B T_\text{t/s} )^{-1}$, where $k_B$ is the Boltzmann constant and $T_\text{t/s}$ the temperature of the tip/sample.
The differential-conductance reads \cite{kroger_surface_2005}:
\begin{equation}
\frac{dI}{dV} (V) = C \int_{-\infty}^{+\infty} n(E) \chi_T (E-V) dE = C n\ast \chi_T.
\end{equation}
\noindent $\chi_T$ describes the broadening due to temperature. It is defined as:
\begin{equation}
\chi_T = \frac{\beta_\text{t}}{4 \left[ \cosh ( \beta_\text{t} V /2 ) \right ]^2}.
\end{equation}

$\chi_T$ is peak-shaped with a full width at half maximum (FWHM) of $3.5 k_B T_\text{t}$
(\SI{1.3}{\milli\volt} for the STM used here at \SI{4.4}{\kelvin}).

\begin{figure}
\begin{center}
\includegraphics{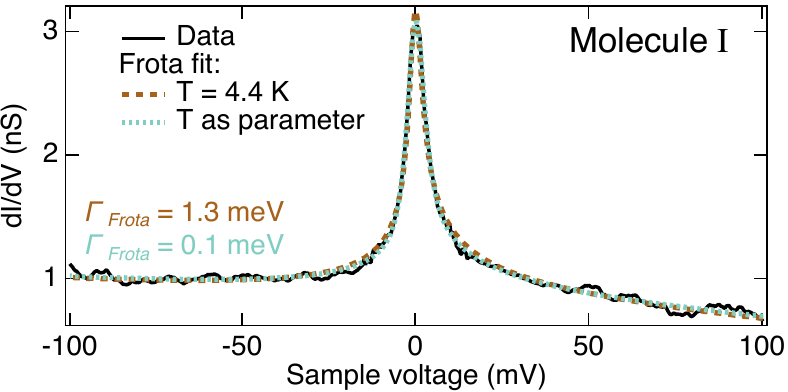}
\end{center}
\caption{
$dI/dV$ spectrum on molecule I along with two fits using temperature-broadened Frota functions.
$T$ was either fixed to \SI{4.4}{\kelvin} (dashed brown) or treated as an adjustable parameter (dashed blue).
The respective $R^2$ are 0.988 and 0.993.
The best fit temperature is \SI{15 \pm 1}{\kelvin}.} 
\label{fig:temperatureEffectOnFit}
\end{figure}

The temperature broadening discussed so far depends on the temperature of the tip.
While the temperature of the sample can usually be measured, the tip temperature may be less well known.
This problem is particularly severe when the temperature of a low temperature STM is varied, because the settling time of the temperature may be rather different for the
tip and other parts the instrument.
An uncertainty of the tip temperature, however, translates into a significant uncertainty of the extracted intrinsic line width ($\Gamma_\text{Frota}$) as demonstrated in Fig.~\ref{fig:temperatureEffectOnFit}.
The dashed brown curve is a fit of the Kondo resonance with a temperature-broadened Frota function, $Frota \ast \chi_T$, where the temperature of the tip is fixed to \SI{4.4}{\kelvin}.
The fit yields $\Gamma_\text{Frota}$ of \SI{1.3\pm 0.1}{\milli\electronvolt}, slightly lower than the value extracted neglecting temperature (\SI{1.5}{\milli\electronvolt}). 
When the tip temperature is treated as an adjustable parameter (Fig.~\ref{fig:temperatureEffectOnFit}, blue dashed curve), $\Gamma_\text{Frota}$ is substantially reduced to \SI{0.1 \pm 0.4}{\milli\electronvolt} (tip temperature for best fit: \SI{15 \pm 1}{\kelvin}).
These results suggest that reliable temperature-dependent measurements of the intrinsic Kondo resonance width require a precise measurement of the tip temperature.

\begin{figure}
\begin{center}
\includegraphics{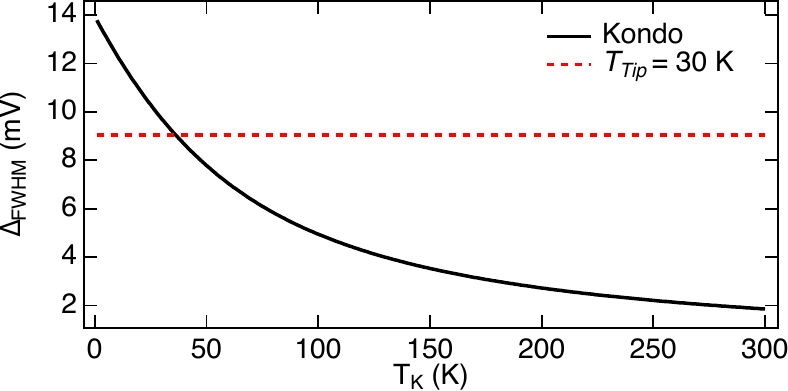}
\end{center}
\caption{Estimated change of the Kondo-resonance FWHM between $T = 0$ and \SI{30}{\kelvin} as a function of the Kondo temperature (black).
The FWHM of the broadening function for a tip temperature of \SI{30}{\kelvin} is indicated by a dashed red line.
For $T_K > \SI{36}{\kelvin}$, the instrumental broadening exceeds the intrinsic width.
For comparison, molecules I and II have estimated Kondo-temperatures $T_K = \SI{35}{\kelvin}$ and \SI{185}{\kelvin}, respectively.} 
\label{fig:expectedBroadeningWithTemperature}
\end{figure}

The intrinsic width of the Kondo resonance evolves as follows \cite{nagaoka_temperature_2002,ternes_spectroscopic_2008,ternes_probing_2017}:
\begin{equation}
FWHM = \sqrt{\left( \alpha k_B T \right)^2 + (2 k_B T_K)^2}.
\label{eq:KondoTDependence}
\end{equation}
\noindent $\alpha$ is a parameter.
Theoretically $\alpha = 2 \pi$ (Ref.~\cite{nagaoka_temperature_2002}) while experimentally a value of 5.4 (Ref.~\cite{otte_role_2008}) was found.
To the best of our knowledge, the discrepancy between the different $\alpha$ has not yet been resolved.
Nonetheless, we have seen that the experimental extraction of the intrinsic FWHM, and henceforth of the $\alpha$, critically depends on the knowledge of the tip temperature and the subsequent deconvolution of the broadening due to temperature.
Note that different forms of Eq.~\ref{eq:KondoTDependence} may be found in the literature, which depend on the definition of $T_K$.
Here we have used the definition $T_K = FWHM /(2 k_B)$.

Temperature-dependent measurements of the Kondo resonance are typically performed from the lowest achieved temperature to approximately \SI{30}{\kelvin}.
Figure~\ref{fig:expectedBroadeningWithTemperature} displays the corresponding intrinsic change of FWHM of a Kondo resonance as a function of the Kondo temperature of the system.
The dashed red line in Fig.~\ref{fig:expectedBroadeningWithTemperature} shows the FWHM of the temperature-broadening function at a tip temperature of \SI{30}{\kelvin}.
The comparison illustrates that temperature-dependent measurements of the intrinsic width are not easy unless $T_K$ is low.

\section{Conclusion}

Measurements on a Kondo system, ReA molecules on Au(111), and model calculations, demonstrate that the experimental line shape of the Kondo resonance is affected by a number of instrumental factors.
As expected, the modulation voltage used for the lock-in detection and noise (at any frequency) on the sample voltage are important sources of broadening.
Interestingly, the effect of broadening is to make a Frota line resemble a Fano line.
As neither of these shapes is an accurate description of the broadened line, the results of fits depend on the voltage range used.

The temperature $T$ of the STM tip adds further broadening.
To still extract a useful estimate of the Kondo temperature, the uncertainty of $T$ must be small, a requirement that is not easily met when the temperature of the instrument is varied.

\ack

Financial support by the Deutsche Forschungsgemeinschaft through SFB 677.
This project has received funding from the European Union's Horizon 2020 research and innovation programme under grant agreement No.~766726.

\section*{References}
\bibliographystyle{iopart-num}
\bibliography{kondoShapeBib}

\end{document}